\documentclass[aps,prb,twocolumn,superscriptaddress,floatfix]{revtex4}
\usepackage{graphicx,epsf}
\usepackage{amsmath}
\usepackage{amssymb}
\usepackage{color}

\begin{document}

\title{
Two-channel pseudogap Kondo and Anderson models: \\
Quantum phase transitions and non-Fermi liquids
}

\author{Imke Schneider}
\affiliation{Institut f\"ur Theoretische Physik, Technische Universit\"at Dresden,
01062 Dresden, Germany}
\author{Lars Fritz}
\affiliation{Institut f\"ur Theoretische Physik, Universit\"at zu K\"oln,
Z\"ulpicher Stra\ss e 77, 50937 K\"oln, Germany}
\author{Frithjof B.~Anders}
\affiliation{Lehrstuhl f\"ur Theoretische Physik II, Technische Universit\"at Dortmund,
Otto-Hahn-Str.~4, 44221 Dortmund, Germany}
\author{Adel Benlagra}
\affiliation{Institut f\"ur Theoretische Physik, Technische Universit\"at Dresden,
01062 Dresden, Germany}
\author{Matthias Vojta}
\affiliation{Institut f\"ur Theoretische Physik, Technische Universit\"at Dresden,
01062 Dresden, Germany}

\date{\today}

\newcommand{\hybb}  {g_0}               
\newcommand{\hyb}   {g}                 
\newcommand{\epsb}  {\varepsilon_0}     
\newcommand{\eps}   {\varepsilon}       
\newcommand{\Jb}    {J_{\rm K}}         
\newcommand{\vvb}   {V_0}               
\newcommand{\vv}    {v}                 
\newcommand{\lam}   {\lambda_0}         
\newcommand{\TK}    {T_{\rm K}}
\newcommand{\Jc}    {J_{c}}
\newcommand{\rmax}  {r_{\rm max}}

\newcommand{\w}     {\omega}

\begin{abstract}
We discuss the two-channel Kondo problem with a pseudogap density of states,
$\rho(\w)\propto|\w|^r$, of the bath fermions. Combining both analytical and numerical
renormalization group techniques, we characterize the impurity phases and quantum phase
transitions of the relevant Kondo and Anderson models. The line of stable points, corresponding to
the overscreened non-Fermi liquid behavior of the metallic $r=0$ case, is replaced by a stable
particle-hole symmetric intermediate-coupling fixed point for $0<r<\rmax\approx0.23$. For
$r>\rmax$, this non-Fermi liquid phase disappears, and instead a critical fixed point with
an emergent spin--channel symmetry appears, controlling the quantum phase transition
between two phases with stable spin and channel moments, respectively. We propose
low-energy field theories to describe the quantum phase transitions, all being formulated
in fermionic variables. We employ epsilon expansion techniques to calculate critical
properties near the critical dimensions $r=0$ and $r=1$, the latter being potentially
relevant for two-channel Kondo impurities in neutral graphene. We find the analytical
results to be in excellent agreement with those obtained from applying Wilson's numerical
renormalization group technique.
\end{abstract}

\maketitle


\section{Introduction}

The two-channel Kondo effect represents a prime example of non-Fermi liquid behavior
arising from a stable intermediate-coupling fixed point.\cite{nozieres1980}
Theoretically, its physics is essentially understood, thanks to an exact solution by
Bethe ansatz.\cite{andrei1984,wiegmann1985}  In addition, boundary conformal field theory
(CFT) has proved to be a powerful technique to study the low-energy properties of the
multi-channel Kondo model\cite{affleck1991} allowing, in particular, the
calculation of exact asymptotic Green's functions.\cite{ludwig1991} CFT techniques have
also been used to calculate exact crossover Green's functions.\cite{sela}
Further, by means of Abelian bosonization and
subsequent re-fermionization  it has been possible to map the two-channel Kondo
problem onto a resonant-level model which reduces to a free fermion form for a
particular value of exchange anisotropy.\cite{emery1992}

On the experimental side, a number of heavy-fermion materials, displaying deviations from
Fermi-liquid behavior, have been speculated to realize two-channel Kondo physics arising
from non-Kramers doublet ground states of U or Pr ions.\cite{coxzawa,cox1987,schiller1998} However, to our
knowledge, there is no unambiguous verification of these proposals.
Consequently, various attempts were made to realize the two-channel Kondo effect in
nano\-structures, and indeed success was reported\cite{potok} for a setup of a semiconductor quantum
dot coupled to two reservoirs.\cite{oreg}
Very recently, signatures of two-channel Kondo behavior of magnetic adatoms on graphene
have been reported,\cite{manoharan} and this motivates to discuss two-channel Kondo
impurities in non-metallic hosts. In particular, neutral graphene realizes a pseudogap
density of states (DOS), $\rho(\omega)\propto|\omega|^r$ with $r=1$, at low energies.

The single-channel pseudogap Kondo problem has been studied extensively in the context of
Kondo impurities in unconventional superconductors. The main difference to the familiar
metallic Kondo problem\cite{hewson} is the absence of screening at small Kondo coupling
$J$, leading to a quantum phase transition upon increasing
$J$.\cite{withoff,cassa,bulla,tolya2,GBI} The universality class of this phase transition
changes as function of $r$,\cite{GBI} and the relevant low-energy field theories have
been worked out in detail in Refs.~\onlinecite{VF04,FV04}.

Although two-channel Kondo physics has been speculated about in the context of
graphene,\cite{baskaran} the two-channel pseudogap Kondo model has received little
attention. A central question is about the fate and character of the non-Fermi-liquid
phase at finite $r$.
To our knowledge, the only study of the model has been reported in a brief
section of Ref.~\onlinecite{GBI}, but there only numerical results were given
for small bath exponents $r$.\cite{GBI_foot}

The purpose of this paper is to close this gap: We shall investigate the two-channel
Kondo and Anderson models with a pseudogap DOS in some detail, using both analytical and
numerical renormalization group (RG) techniques.
Our main findings for the two-channel Kondo model are:

(A) The overscreened non-Fermi liquid (NFL) phase of the metallic two-channel Kondo model
\cite{hewson} survives for $0< r < \rmax \approx 0.23$, albeit with an important
modification: It is no longer represented by a {\em line} of NFL fixed points (where
particle--hole (p-h) asymmetry is marginal), but instead there is only an isolated stable
p-h symmetric NFL fixed point, i.e., p-h asymmetry is irrelevant for
$r>0$.\cite{GBI_foot} Furthermore, this stable NFL phase is only reached for Kondo
couplings larger than a critical coupling, i.e., a boundary quantum phase transition
emerges between a local-moment phase with an unscreened spin moment and the NFL phase. In
contrast, for $r > \rmax$ the NFL fixed point disappears, leaving only two stable phases
with unscreened spin or channel (i.e.~flavor) moment, respectively, which are separated
by a quantum phase transition.

(B) The two-channel pseudogap Kondo physics for both $r\lesssim 1$ and $r\geq 1$ can be fully understood
in the language of the two-channel Anderson model, by virtue of a generalization of
the approach presented in Ref.~\onlinecite{FV04}.
The low-energy field theory describing the quantum phase transition between the phases
with free spin and flavor moments is given by a level crossing of a spin doublet and a
flavor doublet minimally coupled to conduction electrons. Similar to the single-channel
pseudogap Kondo problem, $r=1$ is found to play the role of an upper-critical dimension,
where the hybridization is marginal. For $r>1$, the transition is a level
crossing with perturbative corrections, whereas a non-trivial critical fixed point
emerges for $r<1$. This fixed point is shown to display an emergent spin--channel Z$_2$
symmetry. As in the single-channel case, none of the quantum phase transitions is
described by a Landau-Ginzburg-Wilson-type theory of a bosonic order parameter, instead
all are ``fermionic'' in nature.

The following subsection gives a more detailed summary of our results.


\subsection{Summary of results}
\label{sec:res}

The two-channel Kondo model with a pseudogap host density of
states can be written as ${\cal H} = {\cal H}_{\rm K} + {\cal H}_{\rm b}$, with
\begin{eqnarray}
\label{km}
{\cal H}_{\rm K} &=& \sum_i \left[
\Jb \vec{S} \cdot c_{i\sigma}^\dagger(0) \vec{\tau}_{\sigma\sigma'} c_{i\sigma'}(0) +
\vvb c_{i\sigma}^\dagger(0) c_{i\sigma}(0) \right] \\
{\cal H}_{\rm b} &=& \sum_i \int_{-\Lambda}^{\Lambda} dk\,|k|^r \,
  k \, c_{k i\sigma}^\dagger c_{k i\sigma}. \nonumber
\end{eqnarray}
Here, we have represented the bath, ${\cal H}_{\rm b}$, by linearly dispersing chiral
fermions $c_{k i\sigma}$, where $i=1,2$ is the channel index. $\vec{S}$ is a spin-1/2
SU(2) spin, $\vec\tau$ is the vector of Pauli matrices, summation over repeated spin
indices $\sigma$ is implied, and $c_{i\sigma}(0) = \int d k |k|^r c_{k i\sigma}$ is the
conduction electron operator at the impurity site. The spectral density of the
$c_{i\sigma}(0)$ fermions follows the power law $|\omega|^r$ below the ultra-violet (UV)
cutoff $\Lambda$; details of the density of states at high energies are irrelevant for
the discussion in this paper. In addition to the Kondo coupling $\Jb$, we have also
included a potential scatterer of strength $\vvb$ at the impurity site which will be used
to tune p-h asymmetry. (An asymmetry of the high-energy part of the DOS would have a net
effect similar to non-zero $\vvb$; for simplicity we will assume in the following that
the DOS is p-h symmetric.)

As we shall show below, a comprehensive analysis requires to consider -- in addition to
the two-channel Kondo model -- the two-channel Anderson model, commonly written as
${\cal H} = {\cal H}_{\rm A} + {\cal H}_{\rm b}$ with\cite{coxzawa}
\begin{eqnarray}
\label{aim}
{\cal H}_{\rm A} &=&
\varepsilon_s \sum_\sigma |\sigma\rangle\langle\sigma| + \varepsilon_q \sum_i |i\rangle\langle i| \\
&+& \hybb \sum_{k\sigma i} (|\sigma\rangle \langle i| c_{k i\sigma} + {\rm H.c.})
+ \vvb \sum_{\sigma i} c_{i\sigma}^\dagger(0) c_{i\sigma}(0). \nonumber
\end{eqnarray}
Here, the isolated impurity has four states, i.e., a spin doublet
$|\sigma\rangle=|\uparrow\rangle,|\downarrow\rangle$ and a channel doublet
$|i\rangle=|1\rangle,|2\rangle$. Their mass difference, $\epsb\equiv
\varepsilon_s-\varepsilon_q$, will play a role as a tuning parameter of the quantum phase
transition.

\begin{figure}[!t]
\epsfxsize=3in
\centerline{\epsffile{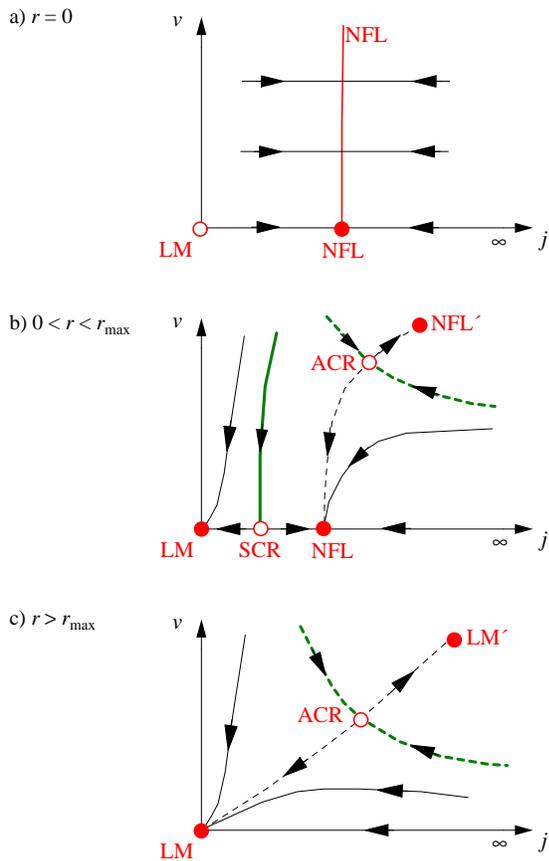}}
\caption{
Schematic RG flow diagrams for the two-channel Kondo model with a pseudogap DOS,
$\rho(\omega)\propto|\omega|^r$. The horizontal axis denotes the renormalized Kondo coupling $j$,
the vertical axis is the renormalized potential scattering $\vv$, representing
particle--hole asymmetry.
Dashed flow lines symbolize a flow out of the plane shown here.
The thick lines correspond to continuous boundary phase
transitions; the full (open) circles are stable (unstable) fixed points.
a) $r\!=\!0$, i.e., the familiar metallic case. For any finite $j$ the flow is towards
the line of NFL fixed points, describing two-channel non-Fermi liquid behavior.
b) $0\!<\!r\!<\!\rmax$:  P-h asymmetry is irrelevant in the NFL phase, such
that the line of fixed points is replaced by a single p-h symmetric NFL fixed
point. The local-moment fixed point LM is stable and separated from NFL by a
p-h symmetric critical SCR fixed point. For $r\to 0$ ($r\to \rmax)$, SCR
approaches LM (NFL). Depending on microscopic details, a second NFL$^\prime$ phase may be reached
at large couplings and asymmetries, separated by a critical ACR fixed point, see text.
c) $r\!\geq\!\rmax$: The NFL phase disappears, and the only phase transition is between
LM and LM$^\prime$, the latter representing a free channel (i.e. flavor) moment. This transition is controlled
by ACR. Note that the character of this transition changes at $r=1$, where ACR merges
with FImp, see text.
}
\label{fig:flow2ck}
\end{figure}

\begin{figure}[!t]
\epsfxsize=3.1in
\centerline{\epsffile{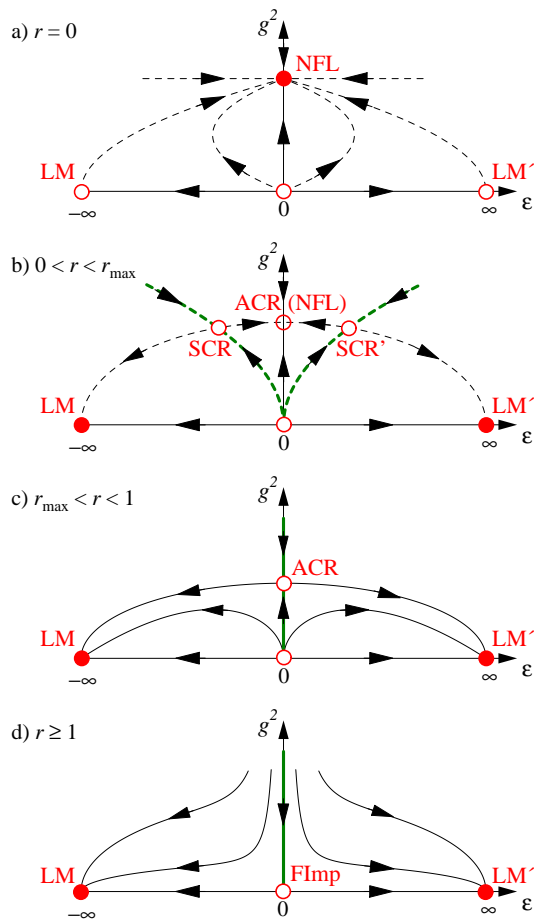}}
\caption{
As Fig.~\ref{fig:flow2ck}, but for the two-channel Anderson model. The horizontal axis
denotes the energy difference $\eps$ between spin and flavor impurity levels, the
vertical axis is the renormalized hybridization $\hyb$. The diagrams represent cuts,
taken at $\vv=0$, through the full RG flow. The flow
diagrams are mirror-symmetric by virtue of the $Z_2$ transformation
(\protect\ref{ph2}).
a) $r\!=\!0$. The line of NFL fixed points,
describing two-channel non-Fermi liquid behavior crosses the $\vv=0$ plane
at $\eps=0$.
b) $0\!<\!r\!<\!\rmax$: The fixed points LM, LM$^\prime$ are stable, corresponding to unscreened
spin and flavor moments, respectively. Within the $\vv=0$ plane, NFL is replaced by ACR,
whereas two isolated p-h symmetric NFL/NFL$^\prime$ fixed points exist outside this plane.
The transitions to NFL/NFL$^\prime$ are controlled by the
p-h symmetric SCR/SCR$^\prime$ fixed points (located outside the $\vv=0$ plane).
c) $\rmax\!\leq\!r\!<\!1$: $\hyb$ is still relevant at $\eps=0$.
However, NFL is now replaced by a single unstable fixed point (ACR) located in the $\vv=0$
plane. At finite $\hyb$, the transition
between the two stable fixed points LM and LM$^\prime$ is controlled by ACR.
d) $r\!\geq\!1$: $\hyb$ is irrelevant, and the only transition is
a level crossing (with perturbative corrections) occurring at $\hyb=\eps=0$, i.e., at the
free-impurity fixed point (FImp).
}
\label{fig:flow2cam}
\end{figure}

For a given value of the bath exponent $r$, the two-channel pseudogap Kondo and Anderson
models display common RG fixed points. The phase diagram and critical behavior depend on
$r$, with $r=0$, $r=\rmax$, and $r=1$ marking qualitative changes and playing the role of
critical ``dimensions''. In the following, we describe our central results for the phase
diagrams and RG flows, which are partially consistent\cite{GBI_foot} with the ones
reported in Ref.~\onlinecite{GBI}.
The qualitative behavior is visualized in the RG flow diagrams in Fig.~\ref{fig:flow2ck}
for the two-channel Kondo model and Fig.~\ref{fig:flow2cam} for the two-channel Anderson
model, respectively. In the latter case, a cut through the RG flow at $\vvb=0$ is shown.

The metallic case $r=0$, has been studied extensively, and a line of infrared stable NFL
fixed points governs the behavior at any finite coupling -- this is the well-known
two-channel (or overscreened) Kondo effect. In the two-channel Anderson model, this line
of fixed points can be accessed by varying $\epsb$, i.e., initial parameters with
different $\epsb$ flow to different fixed points along this line\cite{andrei02} -- note
that this flow leaves the $\vv=0$ plane for $\eps \neq 0$ (dashed in
Fig.~\ref{fig:flow2cam}, all symbols denote the renormalized coupling parameter).

For positive $r$ with $0<r<\rmax$, the line of stable NFL fixed points collapses to an
isolated p-h symmetric NFL fixed point. In addition, the local-moment fixed point (LM) of
an unscreened spin moment now becomes stable. In the language of the Kondo model, LM
corresponds to $j=\vv=0$, while in the Anderson model it corresponds to $\eps=-\infty$,
$\hyb=0$. The phase
transition between LM and NFL is controlled by a critical p-h symmetric fixed point
(SCR); note that this p-h symmetric fixed point is located outside the $\vv=0$ plane
for the Anderson model shown in Fig.~\ref{fig:flow2cam}. As $r\rightarrow 0$ SCR
approaches LM and the critical behavior of SCR is perturbatively accessible for small
Kondo coupling $J_{\rm K}$. The phase diagram of the Anderson model is mirror symmetric,
i.e., there exists also an unscreened channel (or flavor) local-moment fixed point LM$^\prime$ at
$\eps=\infty$, $\hyb=0$ \cite{asc_note} and a corresponding critical
intermediate-coupling fixed point SCR$^\prime$ at positive $\eps$.

As $r\rightarrow \rmax$ SCR approaches NFL, and the two fixed points disappear for
$r>\rmax$. In the p-h symmetric Kondo model, this implies that the flow is towards LM for any
value of $j$, but for large asymmetries, LM$^\prime$ may be reached.\cite{asc_note}
In the Anderson model, the hybridization $\hyb$ remains relevant at $\eps=0$
for $\rmax<r<1$, but the flow is towards a single unstable intermediate coupling fixed
point (ACR) in the $\vv=0$ plane, i.e., ACR is p-h asymmetric, but is invariant under the
$Z_2$ transformation (\protect\ref{ph2}). At finite coupling, the transition between the
two stable fixed points LM and LM$^\prime$ is controlled by ACR.

Finally, as $r\rightarrow 1$ ACR moves toward $\hyb\rightarrow 0$ and for $r\geq 1$ the
phase transition becomes a level crossing with perturbative corrections, controlled by
the free impurity fixed point (FImp) at $\eps=0,\hyb=0$.

\subsection{Outline}

The bulk of this paper is organized as follows:
We start in Sec.~\ref{sec:model} by discussing the relevant impurity models, suitably generalized to
higher degeneracies, along with their underlying symmetries.
In Sec.~\ref{sec:nrg} we present selected results from Wilson's numerical renormalization
group (NRG) for the two-channel pseudogap Anderson and Kondo models which illustrate the
content of the flow diagrams in Figs.~\ref{fig:flow2ck} and \ref{fig:flow2cam}. In
particular, we show properties of the non-trivial intermediate-coupling fixed points as
function of the bath exponent $r$.
Secs.~\ref{sec:weak} and \ref{sec:sym} are devoted to the epsilon expansion studies of
the critical fixed points, using the variables of the Kondo model
(Sec.~\ref{sec:weak}) and that of the Anderson model (Sec.~\ref{sec:sym}). The latter
provide access to the physics near the upper-critical dimension $r=1$.
Concluding remarks will close the paper.
A discussion of the Majorana representation of the two-channel Kondo problem and its fate
in the presence of a pseudogap DOS is relegated to the appendix.


\section{Models, symmetries, and mappings}
\label{sec:model}

\subsection{Anderson model}

The two-channel Anderson model can be understood as describing the level crossing between
two impurity doublets -- one spin doublet and one channel (i.e.~flavor) doublet -- coupled to
conduction electrons via a hybridization term. The model features an
SU(2)$_{\rm spin}$ $\times$ SU(2)$_{\rm flavor}$ symmetry. This can be straightforwardly generalized to
SU($N$)$_{\rm spin}$ $\times$ SU($K$)$_{\rm flavor}$ symmetry, where $N$ is the number
of spin degrees of freedom and $K$ the number of flavors. The Hamiltonian can be
written as:
\begin{eqnarray}
\label{aimNK}
{\cal H}_{\rm A} &=&{\cal H}_{\rm b}+
\varepsilon_s \sum_\sigma |\sigma\rangle\langle\sigma| + \varepsilon_q \sum_\alpha |\bar{\alpha}\rangle\langle \bar{\alpha}| \\
&+& \hybb \sum_{k \alpha\sigma } (|\sigma\rangle \langle \bar{\alpha}| c_{k \alpha \sigma} + {\rm H.c.})
+\vvb \sum_{\alpha \sigma } c_{\alpha\sigma}^\dagger(0) c_{\alpha\sigma}(0)\nonumber
\end{eqnarray}
Here, the conduction electrons $c_{k \alpha \sigma}$ transform under a fundamental
representation of SU$(N)$ and SU$(K)$ and carry the corresponding spin $\sigma$ and
flavor $\alpha$ indices. $\bar{\alpha}$ indicates a transformation behavior according to
the conjugate representation.
For $N=2$, $K=1$, the Hamiltonian in Eq.~\eqref{aimNK} describes the single-channel
Anderson model in the limit of infinite Coulomb repulsion ($U=\infty$), studied using RG in
Ref.~\onlinecite{FV04}. In case of a metallic host, $r=0$, the multi-channel Anderson
model is integrable and has been solved  using the Bethe
Ansatz \cite{bolech2005,andrei02,andrei03} and the numerical RG method.\cite{anders2005}

The metallic two-channel Anderson model, i.e.~$N=2$, $K=2$, has been proposed as a model
for the observed non-Fermi liquid behavior of the heavy-fermion superconductor
UBe$_{13}$.\cite{stewart1983} In this scenario, the 5f$^2$ ground state of the U ion is
identified as the $\Gamma_3$ non-magnetic quadrupolar doublet, while the first excited
state is the 5f$^3$ $\Gamma_7$ magnetic doublet.\cite{coxzawa} This then can promote a
quadrupolar Kondo effect where the quadrupolar doublet is quenched by the hybridization
with $\Gamma_8$ conduction electrons which carry both magnetic and quadrupolar degrees of
freedom.\cite{cox1987} In particular, since the energy difference between the two
doublets appears to be small, a mixed valence state is likely requiring the study of the
full Anderson model.\cite{aliev1995}
Consequently, the model \eqref{aimNK} with a pseudogap DOS is of potential relevance not
only to two-channel impurities in graphene, but also to quadrupolar Kondo impurities in
unconventional superconductors.

The Anderson model \eqref{aimNK} is {\em not} particle--hole symmetric for any value of
$\vvb$, due to the asymmetric structure of the impurity. However, p-h symmetry
is dynamically restored for $0<r<\rmax$ both inside the NFL/NFL$^\prime$ phases and at the
critical SCR/SCR$^\prime$ fixed points, see Fig.~\ref{fig:flow2ck}.

Interestingly, for $N=K$ and a p-h symmetric bath, the Anderson model displays a
spin--channel symmetry, i.e., is invariant under the combined transformation
\begin{eqnarray}
|\sigma \rangle &\leftrightarrow& |\bar{\alpha}\rangle \nonumber\\
c_{k \alpha \sigma} &\leftrightarrow& c_{k \alpha \sigma}^\dagger \nonumber\\
\epsb &\leftrightarrow& -\epsb \nonumber\\
\vvb  &\leftrightarrow& -\vvb
\label{ph2}
\end{eqnarray}
Here, the spin-carrying impurity states are transformed
into the flavor-carrying states and vice versa, i.e.,
the two SU($N$) sectors are interchanged, together with a p-h transformation.

\subsection{Kondo models}

The Anderson model \eqref{aimNK} has two Kondo limits. On the one hand, for $\epsb =
\varepsilon_s-\varepsilon_q \to-\infty$ it maps to a  $K$-channel SU($N$)$_{\rm
spin}$-Kondo model, where a spinful impurity is coupled to $K$ channels of conduction
electrons. For $N=2$ the Hamiltonian reads
\begin{eqnarray}
\label{kmK}
{\cal H}_{\rm K} ={ \cal H}_{\rm b}&+&
\Jb \vec{S} \cdot \sum_{\alpha \sigma \sigma'} c_{\alpha\sigma}^\dagger(0) \vec{\tau}_{\sigma\sigma'} c_{\alpha\sigma'}(0) \\
&+&
\vvb \sum_{\alpha \sigma } c_{\alpha\sigma}^\dagger(0) c_{\alpha\sigma}(0)\nonumber
\end{eqnarray}
where $\vec{S}$ is a spin-1/2 SU(2) spin and $\sigma,\sigma'=\uparrow,\downarrow$.
For $N>2$ the impurity spin is in a fundamental representation of SU($N$).
The parameters of the Kondo model \eqref{kmK} are related to that of the $\vvb=0$
Anderson model (\ref{aimNK}) through:
\begin{equation}
\Jb = N \vvb = \frac {\hybb^2} {|\epsb|}\,.
\label{j-swo}
\end{equation}
The Kondo limit is reached by taking $\epsb\to -\infty$, $\hybb\to\infty$, keeping
$\Jb$ fixed. Note that a potential scattering term is always generated.

On the other hand, for $\epsb\to+\infty$ the Anderson model can be mapped to a
$N$-channel SU($K$)$_{\rm flavor}$-Kondo model, where $\vec{S}$ represents a SU($K$)
impurity which is screened by the $N$ spin degrees of freedom of the conduction
electrons. Such multi-channel flavor Kondo effect is relevant, e.g., to the charging
process of a quantum box, where the flavor degree of freedom is taken
by the physical charge.\cite{matveev,LSA03}

We note that the multi-channel Kondo model cannot be obtained by a Schrieffer-Wolff
transformation from any standard Anderson model (i.e., written with free-electron
operators and local Coulomb interaction).

\subsection{Large-$N$ limit}

The SU($N$) multi-channel Kondo model can be solved in a dynamic large-$N$ limit for both
fully symmetric (bosonic) and fully antisymmetric (fermionic) representations of
SU($N$).\cite{page} The fermionic version of this solution, with $K=\gamma N$ and
$K,N\to\infty$, has been generalized to the pseudogap case.\cite{mv01} The large-$N$ phase
diagram, Fig.~1 of Ref.~\onlinecite{mv01}, is similar to that of the $N=K=2$ case, i.e.,
the overscreened non-Fermi liquid phase survives for small $r$, where it is reached for a
certain range of couplings only, while this phase disappears for larger $r\lesssim 1$. Also, all leading
anomalous dimensions vanish for $r>1$.

Two qualitative differences between the large-$N$ scenario and $N=K=2$ are worth noting:
(i) The critical ``dimension'' $\rmax$ of $N=K=2$ splits into two in the large-$N$
limit, with their values and the detailed behavior depending upon the value of $\gamma$.
(ii) The quantum phase transitions in the large-$N$ limit are governed by {\em lines} of
fixed points, with continuously varying exponents as function of the particle--hole
asymmetry, in contrast to the isolated critical fixed points SCR and ACR in
Figs.~\ref{fig:flow2ck} and \ref{fig:flow2cam}. This implies that the scaling dimension
of $\vvb$ vanishes at criticality as $N\to\infty$, rendering the large-$N$ limit
partially singular. We therefore refrain from a detailed discussion of the models
\eqref{aimNK} and \eqref{kmK} for large $N$.

\subsection{Observables}

In the bulk of the paper, we will focus on a few important observables which characterize
the phases and phase transitions of the impurity models under consideration. Those
include the correlation-length exponent, the impurity entropy, various susceptibilities, and
the conduction-electron T matrix (or impurity spectral function). Their definition is
standard, and we refer the reader to Refs.~\onlinecite{GBI,vbs,mvrev,MKMV,FV04} for a detailed
exposition. Here we only summarize a few key aspects.

Spin susceptibilities, $\chi^{\rm (spin)}$, are obtained by coupling external magnetic
fields both  to the bulk and impurity degrees of freedom as explained in detail in
Ref.~\onlinecite{MKMV}. For the impurity part, here, this reads
\begin{eqnarray}
-H_{\rm{imp}, i} \,\, |\sigma\rangle \lambda^i_{\sigma \sigma^\prime} \langle \sigma^\prime|,
\end{eqnarray}
where $H_{\rm{imp}, i}$ is the magnetic field coupling to the
impurity spin, while the $\lambda^i_{\sigma \sigma^\prime}$ with $i=1,N-1$ are  generators
of SU($N$). In the following, we exploit the SU($N$) symmetry and only evaluate the
corresponding susceptibility tensor in the $1$-direction choosing the representation
$\lambda^1_{\sigma,\sigma^\prime}=\frac{1}{2}(\delta_{\sigma,1}\delta_{\sigma^\prime,1}-\delta_{\sigma,2}\delta_{\sigma^\prime,2})$.\cite{georgi_book}
We proceed as usual by calculating the magnetic susceptibilities via the corresponding
linear response functions. Note that the impurity susceptibility is composed of
\begin{eqnarray}
\chi_{\rm imp}(T)=\chi_{\rm imp,imp}+2\chi_{\rm b,imp}+(\chi_{\rm b,b}-\chi_{\rm b,b}^{0}),
\end{eqnarray}
where $\chi_{\rm imp,imp}$ is the response to $H_{\rm imp}$, $\chi_{\rm b,b}$  measures
the bulk response to the field applied to the bulk, $\chi_{\rm b,imp}$ are the cross
terms, and $\chi_{\rm b,b}^{0}$ denotes the bulk response in the absence of the impurity.
Flavor susceptibilities, $\chi^{\rm (flavor)}$, can be defined in the Anderson model in
analogy to the spin susceptibilities (i.e., with $\sigma\rightarrow\alpha$).

Owing to symmetries, the total magnetization in both the spin and flavor sectors is
conserved. This implies that the impurity contributions to the spin and flavor
susceptibilities, $\chi_{\rm imp}^{\rm (spin)}$ and $\chi_{\rm imp}^{\rm (flavor)}$, do
not acquire anomalous exponents at the intermediate-coupling fixed points, but instead
obey Curie laws with (in general) fractional prefactors. In contrast, the local spin and
flavor susceptibilities follow anomalous power laws, $\chi_{\rm loc}^{\rm (spin)}
\propto T^{-1+\eta_\chi^{\rm (spin)}}$ and $\chi_{\rm loc}^{\rm (flavor)} \propto
T^{-1+\eta_\chi^{\rm (flavor)}}$, with universal $r$-dependent exponents $\eta_\chi$. We
note that a direct calculation of both susceptibilities is only possible in the Anderson
model, as the Kondo limit suppresses the local piece of one of the susceptibilities. To
shorten notation, we employ the convention $\chi \equiv \chi^{\rm (spin)}$ and $\eta_\chi
\equiv \eta_\chi^{\rm (spin)}$ in the following.

Similar to $T\chi_{\rm imp}^{\rm (spin,flavor)}$, the impurity entropy approaches a
universal fractional value as $T\to 0$. The conduction-electron T matrix, on the other
hand, follows an anomalous power law similar to the local susceptibility, $T(\w)
\propto \w^{-1+\eta_T}$.

At the non-Fermi-liquid fixed point of the familiar metallic two-channel Kondo model ($r=0$), power
laws are replaced by logs, $\chi_{\rm loc},\chi_{\rm imp} \propto \ln 1/T$ -- this also
implies that the prefactor of the leading Curie term in $\chi_{\rm imp}$ vanishes due to
an exact compensation.


\section{Selected numerical results}
\label{sec:nrg}

The NRG technique\cite{NRGrev} is ideally suited to study properties of quantum impurity
models, including non-Fermi liquid phases and quantum phase transitions.
Initial NRG results for the two-channel pseudogap Kondo model were shown in
Ref.~\onlinecite{GBI}. Here we extend and complement this early analysis by NRG results
for the two-channel ($N=K=2$) Anderson model. We perform explicit calculations for a bath
density of states $\rho(\w) = (1+r)/(2D) |\w/D|^r \Theta(D^2-\w^2)$ with $D=1$.
Unless otherwise noted, we employ NRG parameters\cite{NRGrev} $\Lambda=6$ and $N_s=600$.

\begin{figure}
\epsfxsize=3.2in
\centerline{\epsffile{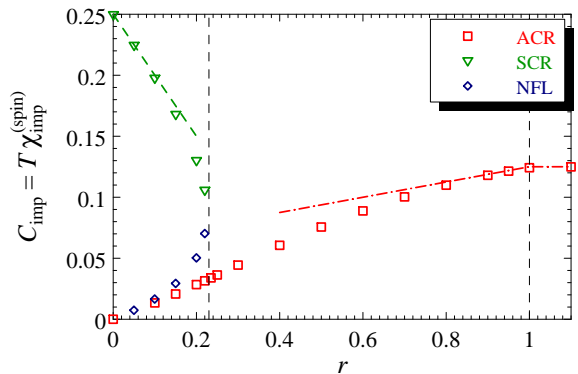}}
\caption{
NRG for the impurity susceptibility, $T\chi_{\rm imp}$, at the
intermediate-coupling fixed points ACR (${\color{red}\square}$), SCR
(${\color{green}\triangle}$), and NFL (${\color{blue}\lozenge}$). Also shown are the
results from the renormalized perturbation theory of the Kondo model, which allows to
access SCR near $r=0$ [Sec.~\protect\ref{sec:weak}, Eq.~(\protect\ref{tchi_weak}),
dashed], and that of the Anderson model, appropriate for ACR for $r\lesssim 1$
[Sec.~\protect\ref{sec:sym}, Eq.~(\protect\ref{tchi_2cam}), dash-dot]. The SCR data have
been partially taken from Ref.~\protect\onlinecite{GBI}. The vertical dashed lines
indicate the critical dimensions $r=\rmax$ and $r=1$.
}
\label{fig:tchi}
\end{figure}

\begin{figure}[b]
\epsfxsize=3.2in
\centerline{\epsffile{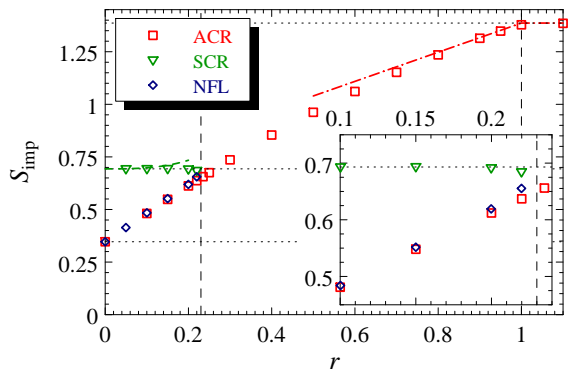}}
\caption{
As Fig. \protect\ref{fig:tchi}, but for the impurity entropy $S_{\rm imp}$.
The perturbative expansions are in
Eq.~(\protect\ref{simp_weak}) (dashed) and
Eq.~(\protect\ref{simp_2cam}) (dash-dot).
The dotted horizontal lines correspond to $S_{\rm imp} = 0.5 \ln 2$, $\ln 2$, and $\ln 4$.
The inset shows a zoom onto the region $r\lesssim \rmax$.
}
\label{fig:simp}
\end{figure}

The qualitative behavior of the two-channel Anderson model is summarized in the flow
diagram in Fig.~\ref{fig:flow2cam}. In addition to the stable LM/LM$^\prime$ phases, these flow
diagrams feature three non-trivial fixed points: the stable NFL/NFL$^\prime$ fixed points and the
critical fixed points SCR/SCR$^\prime$ and ACR.
Some of their key properties are summarized in Figs.~\ref{fig:tchi} and \ref{fig:simp},
which show the numerically determined impurity contributions to the
spin susceptibility and the entropy, respectively, together with analytical results
obtained from the epsilon expansion of Secs.~\ref{sec:weak} and \ref{sec:sym}.
These plots nicely show that NFL and SCR approach each other as $r\to\rmax$, while ACR
evolves continuously near $\rmax$. The fixed-point properties also show that SCR
approaches LM as $r\to 0$, with $T\chi_{\rm imp}\to 1/4$ and $S_{\rm imp}\to \ln 2$, and
ACR approaches FImp as $r\to 1$, with $T\chi_{\rm imp}\to 1/8$ and $S_{\rm imp}\to \ln
4$. Further, the stable NFL fixed point follows $T\chi_{\rm imp}\to 0$ and $S_{\rm imp}\to 0.5\ln
2$ as $r\to 0$ -- the well-known properties of the metallic two-channel Kondo problem.

\begin{figure}[t]
\epsfxsize=3.2in
\centerline{\epsffile{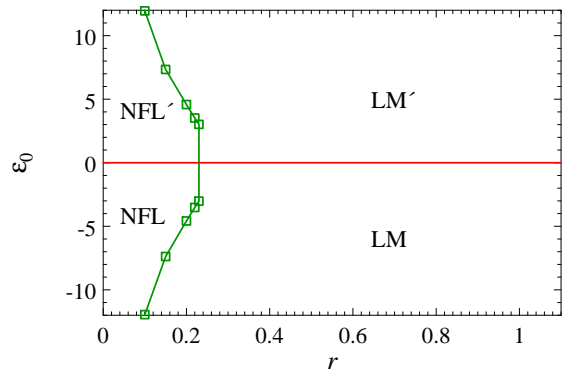}}
\caption{
Phase diagram of the two-channel Anderson model as function of the energy difference
$\epsb$ and the bath exponent $r$, keeping $\hybb^2=4$ 
fixed. The discontinuous change at $r=\rmax$ is apparent.
}
\label{fig:phd}
\end{figure}

The disappearance of both NFL and SCR upon increasing $r$ beyond $\rmax$
implies a {\em discontinuous} evolution of the phase diagram as function of $r$. In
Fig.~\ref{fig:phd} we present a cut through the phase diagram of the Anderson model at
fixed $\hybb$ which illustrates this fact. We note that such a discontinuous evolution
occurs if a stable intermediate-coupling fixed point disappears (here NFL); in contrast,
if a trivial fixed point changes its nature from stable to unstable, the evolution is
continuous, like in the single-channel pseudogap Kondo model.

\begin{figure}[b]
\epsfxsize=3.2in
\centerline{\epsffile{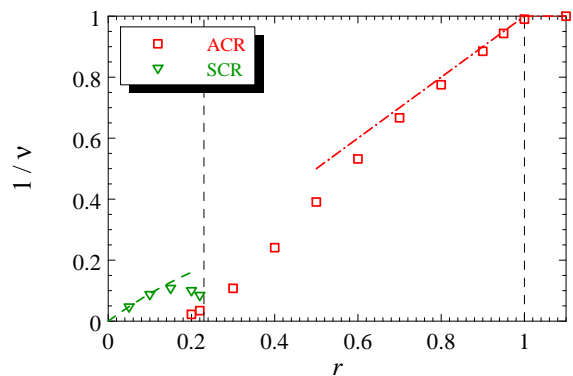}}
\caption{
Inverse correlation-length exponent $1/\nu$ obtained from NRG for the ACR
(${\color{red}\square}$) and SCR (${\color{green}\triangle}$) critical points, together
with the analytical RG results from the expansions in $r$ [Sec.~\protect\ref{sec:weak},
Eq.~(\protect\ref{nuz_weak}), dashed] and in $(1-r)$ [Sec.~\protect\ref{sec:sym},
Eq.~(\protect\ref{nuz_2cam}), dash-dot]. Note that $\nu$ of ACR becomes very large for
small $r$ -- this corresponds to the extremely slow flow from ACR to NFL for $0<r<\rmax$.
}
\label{fig:nuz}
\end{figure}

The correlation-length exponents $\nu$ are displayed in Fig.~\ref{fig:nuz}, illustrating
that $r=0$ and $\rmax$ play the role of lower-critical dimensions for the p-h symmetric
transition controlled by SCR, with $\nu\to\infty$, whereas $r=1$ is the upper-critical
dimension of the ACR transition, with $\nu=1$ for all $r>1$ (and logarithmic corrections
at $r=1$).

\begin{figure}[t]
\epsfxsize=3.2in
\centerline{\epsffile{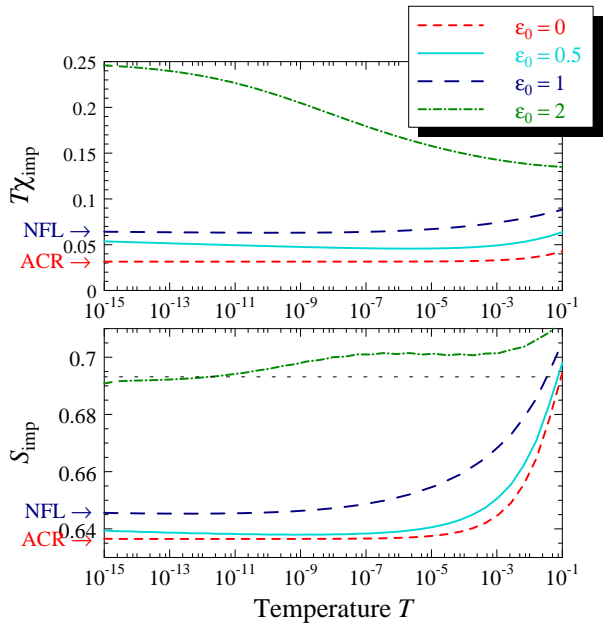}}
\caption{
NRG results for a) the impurity susceptibility $T\chi_{\rm imp}$ and b) the impurity
entropy $S_{\rm imp}$ of the two-channel Anderson model at $r=0.22$. Parameter values are
$\hybb^2=2$ and $\epsb=0$ (dashed), 0.5 (solid), 1 (long-dash), 2 (dash-dot). Both
$\epsb=0.5$ and 1 are located in the NFL phase while $\epsb=2$ corresponds to the
LM phase.
The $\epsb=0.5$ curves show the flow from ACR to NFL upon lowering $T$, where in particular
the impurity entropy {\em increases} along the RG flow. (The flow is too slow to reach
its fixed point within the accessible temperature range.)
}
\label{fig:entr}
\end{figure}

Fig.~\ref{fig:entr} shows the temperature evolution of both $T\chi_{\rm imp}$ and $S_{\rm
imp}$ for $r=0.22$, i.e., slightly below $\rmax$. Here, the flow from ACR to NFL
(compare Fig.~\ref{fig:flow2ck}b) is nicely visible at $\epsb=0.5$ where both $T\chi_{\rm imp}$
and $S_{\rm imp}$ increase along the RG flow. (The large value of $\nu$ at ACR renders
the flow very slow.)
Remarkably, the fact $S_{\rm ACR} < S_{\rm
NFL}$ violates so-called $g$-theorem,\cite{gtheorem} which states that the impurity
entropy should decrease along the flow. As this theorem applies to conformally invariant
systems only, we conclude that the fixed points under consideration are not described by
a conformally invariant theory. (The same conclusion can be drawn for the quantum phase
transitions of the single-channel pseudogap Kondo problem, see Ref.~\onlinecite{FV04}.
Also, such ``uphill flow'' may occur in models with long-range interactions, see e.g.
Ref.~\onlinecite{uphill}.)

\section{Weak-coupling expansion for the multi-channel Kondo model}
\label{sec:weak}

In this section we review the standard weak-coupling expansion\cite{poor} for the multi-channel
SU($2$)-Kondo model in Eq.~(\ref{kmK}), extended to a pseudogapped bath density of
states.\cite{withoff,GBI} As we show below, this expansion captures the properties of the
critical fixed point SCR at small $r$. In principle, it also allows to access the stable
NFL fixed point, but this requires a particular limit of large $K$ which does not allow
to extract quantitative results for the $K=2$ case of interest.

RG equations can be derived, e.g., using the field-theoretic scheme\cite{bgz,lars_book} where logarithmic
divergencies, occurring for $r=0$, are replaced by poles in $r$ by means of
``dimensional'' regularization. Doing so, $r$ will only enter in the bare scaling
dimension of the couplings.
To two-loop order, the equations for the renormalized couplings $j$ and $\vv$ read
\begin{eqnarray}
\label{betaj}
\beta(j) &=& r j - j^2 + \frac{K}{2} j^3,\\
\beta(\vv) &=& r \vv, \nonumber
\end{eqnarray}
where $K$ is the number of equivalent screening channels. Importantly, there is no
renormalization of $\vv$, a result which persists to higher orders.

Apart from the LM fixed point, $j=\vv=0$, the function $\beta(j)$ in Eq.~\eqref{betaj} yields two further
zeros, given by $j = (1\pm\sqrt{1-2Kr})/K$. The smaller one corresponds to an infrared
unstable fixed point at
\begin{equation}
\label{jfix}
j^\ast = r + \frac{K}{2} r^2 + {\cal O}(r^3),~
\vv^\ast = 0
\end{equation}
which can be perturbatively controlled for $r\to 0$ and any $K$. We label this fixed point by SCR. As $r\to 0$ SCR approaches LM.
The larger zero predicts an infrared stable fixed point at
\begin{equation}
\label{jfix2}
j^\ast = \frac{2}{K} - r - \frac{K}{2} r^2 + {\cal O}(r^3),~
\vv^\ast = 0.
\end{equation}
Strictly speaking, this fixed point is perturbatively accessible only if the limits $r\to
0$ and $K\to\infty$ are either taken in this order (this corresponds to $r=0$) or
together such that $Kr$ is kept fixed.
For $r=0$, where $\vv$ is marginal, this zero of $\beta(j)$ is commonly
associated with the line of stable non-Fermi liquid fixed points of the multi-channel Kondo
model, which exists for all $K\geq 2$.
For $r>0$ now $\vv$ becomes irrelevant, in agreement with our numerical results\cite{GBI_foot} which
show that the NFL line of fixed points shrinks to a single p-h symmetric NFL fixed point,
Fig.\ref{fig:flow2ck}.

\subsection{Observables near criticality}
\label{sec:weakobs}

The properties of the p-h symmetric critical fixed point SCR, existing for
$0<r<\rmax$, can be determined in analogy to the single-channel case, with explicit
calculations  given e.g. in Ref.~\onlinecite{MKMV}.
Expanding the beta function (\ref{betaj}) around the fixed point value (\ref{jfix})
yields the correlation length exponent $\nu$:
\begin{equation}
\frac{1}{\nu} = r - \frac{K}{2} r^2 +  {\cal O}(r^3)\,.
\label{nuz_weak}
\end{equation}

The leading-order perturbative corrections to the impurity susceptibility and entropy are given by
\begin{eqnarray}
\Delta (T\chi_{\rm imp}) &=& - \frac{Kj}{4}\,,\nonumber\\
\Delta S_{\rm imp}       &=& \frac{3\pi^2 \ln 2}{8} \, Kj^2 \, r .
\end{eqnarray}
Inserting the fixed-point value $j^\ast$ \eqref{jfix} we obtain
\begin{eqnarray}
\label{tchi_weak}
T\chi_{\rm imp} &=& \frac{1}{4} (1 - Kr) + {\cal O}(r^2)\,, \\
\label{simp_weak}
S_{\rm imp} &=& \ln 2 \bigg(1 + \frac{3\pi^2}{8}\,K r^3\bigg) + {\cal O}(r^5) \,.
\end{eqnarray}
The anomalous exponent of the local susceptibility is given by $\eta_\chi = K j^2$ to
leading order, which evaluates to
\begin{equation}
\label{etachi_weak}
\eta_\chi = K r^2 + {\cal O}(r^3) \,.
\end{equation}
Finally, the T matrix exponent is $\eta_T = 1-j$ (with no factor of $K$, as the T matrix
describes the scattering of electrons from one specific channel), resulting in
\begin{equation}
\label{etaT_weak}
\eta_T = 1 - r.
\end{equation}
Note that this result is exact.\cite{FV04}


\section{Hybridization expansion for the multi-channel Anderson model}
\label{sec:sym}

We now turn our attention to the multi-channel Anderson model. As we show below, the
variables of the Anderson model will allow us to obtain an essentially complete
understanding of the multi-channel pseudogap Kondo effect both for $r\lesssim 1$ and $r\geq
1$. A similar conclusion was reached for the single-channel case in
Refs.~\onlinecite{VF04,FV04}, and -- on a technical level -- our work represents a
generalization of the calculation for the infinite-$U$ Anderson model in Ref.~\onlinecite{FV04}.

\subsection{Trivial fixed points}

For vanishing hybridization $\hybb$, the multi-channel Anderson model (\ref{aimNK})
features three trivial fixed points: for $\epsb<0$ the ground state is the spinful
$N$-fold degenerate local-moment state (LM) and, analogously, for $\epsb>0$ it is the
$K$-fold degenerate flavor local-moment state (LM$^\prime$). In these cases the impurity
entropy equals $\ln N$ and $\ln K$, respectively. For $\epsb=0$ there are $(N+K)$
degenerate impurity states, we refer to this
 as the free-impurity fixed point (FImp), with entropy $\ln(N+K)$.
The impurity spin susceptibilities are
\begin{equation}
T\chi_{\rm imp}^{\rm (spin)} = \left\{
\begin{array}{ll}
\frac{1}{2N}     & \mbox{LM} \\
\frac{1}{2(N+K)} & \mbox{FImp} \\
0                & \mbox{LM}^\prime \\
\end{array}
\right. .
\end{equation}
The corresponding values of $T\chi_{\rm imp}^{\rm (flavor)}$ follow via Eq.~\eqref{ph2}
from LM$\,\leftrightarrow\,$LM$^\prime$.
The hybridization term, $\hybb$, is irrelevant at LM and LM$^\prime$ for $r>0$.

\subsection{Hybridization expansion and upper critical dimension}
\label{sec:rghyb}

In the following we perform an expansion around the FImp fixed point, i.e.,
around $\epsb=0$, $\hybb=0$.

The impurity states are represented by bosonic operators $b_{\bar{\alpha}}^\dagger$ for
$\alpha = 1,\ldots,K$ and fermionic operators $f_\sigma$ for $\sigma=1,\ldots,N$.
Single occupancy of the localized levels is enforced by the Hilbert space constraint
$\hat{Q}\equiv\sum_{\alpha}b_{\bar{\alpha}}^\dagger b_{\bar{\alpha}}+\sum_\sigma
f_\sigma^\dagger f_\sigma=1$ which will be implemented using a chemical potential
$\lambda_0\rightarrow \infty$. Observables are then calculated as \cite{lambda,costi}
\begin{eqnarray}
\label{observables_lambda}
\langle \hat{O}\rangle=\lim_{\lambda_0\rightarrow \infty}\frac{\langle\hat{Q}\hat{O}\rangle_{\lambda_0}}{\langle\hat{Q}\rangle_{\lambda_0}},
\end{eqnarray}
where $\langle \cdots\rangle_{\lambda_0}$ denotes the thermal expectation value in the
presence of the chemical potential $\lambda_0$.

Furthermore, we need to introduce chemical-potential counter-terms
which cancel the shift of the critical point occurring
in perturbation theory upon taking the limit of infinite UV cutoff.
Technically, this shift arises from the real parts
of the self-energies of the $b_{\bar{\alpha}}$ and $f_\sigma$ particles.
We introduce the counter-terms as additional chemical
potential for the auxiliary particles,
\begin{equation}
\delta\lambda_b \, b_{\bar{\alpha}}^\dagger b_{\bar{\alpha}} \,,~
\delta\lambda_f \, f_\sigma^\dagger f_\sigma \,.
\label{counter}
\end{equation}
The $\delta\lambda_{b,f}$ have to be determined order by order in an expansion in
$\hybb$.
Note that counter-term contributions in observables in general enter both
numerator and denominator in Eq.~(\ref{observables_lambda}).

In the
path integral form the model \eqref{aimNK} is written as
\begin{eqnarray}
\label{th}
{\cal S} = \int_0^\beta d \tau
&\bigg[&
   \bar{f}_\sigma (\partial_\tau + \lam + \varepsilon_s+\delta\lambda_f ) f_\sigma \nonumber\\
  &+& \bar{b}_{\bar{\alpha}} (\partial_\tau + \lam + \varepsilon_q+\delta\lambda_b ) b_{\bar{\alpha}} \nonumber\\
  &+& \hybb \left(\bar{f}_\sigma b_{\bar{\alpha}} c_{\alpha\sigma}(0) + {\rm c.c.}\right) \nonumber\\
  &+& \int_{-\Lambda}^{\Lambda} d k\,|k|^r \,
  {\bar c}_{k \alpha\sigma} (\partial_\tau+k) c_{k \alpha \sigma}
\bigg],
\end{eqnarray}
where $\lam$ is the chemical potential enforcing the constraint exactly. Here, we
implicitly sum over $\sigma$ and $\alpha$.

The model (\ref{th}) shows a transition driven by variation of
$\epsb=\varepsilon_s-\varepsilon_q$. At the $\epsb=\hybb=0$ fixed point,
tree level scaling analysis shows that
\begin{eqnarray}
{\rm dim}[\hybb] &=& \frac{1-r}{2} \equiv {\bar r} \,, \\
{\rm dim}[\epsb] &=& 1 \,. \nonumber
\end{eqnarray}
As in the single-channel model,\cite{VF04,FV04} this establishes the role of $r=1$ as an
upper-critical dimension where $\hybb$ is marginal.

We perform a field-theoretical RG analysis using the minimal subtraction scheme.\cite{bgz,lars_book}
Renormalized fields and dimensionless couplings are introduced according to
\begin{eqnarray}
f_\sigma&=&\sqrt{Z_f} f_{R\sigma},\\
b_{\bar{\alpha}}&=&\sqrt{Z_b} b_{R\bar{\alpha}},\\
\label{g_r}
\hybb &=&\frac{\mu^{\bar{r}}Z_g}{\sqrt{Z_fZ_b}} \hyb,
\end{eqnarray}
where $\mu$ is the renormalization group energy scale. 
No renormalizations are needed for the bulk fermions as their self interaction is assumed
to be irrelevant. The RG is conveniently performed at criticality, i.e. we assume that
$\epsb$ is tuned to the critical line and set $\varepsilon_q=\varepsilon_s=0$.

\begin{figure}
\epsfxsize=3.2in
\centerline{\epsffile{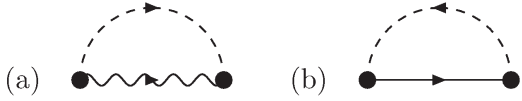}}
\caption{
Feynman diagrams entering the self energies up to quadratic order in $\hybb$.
Full/dashed/wiggly lines denote $f_\sigma/c_\sigma/b_{\bar{\alpha}}$ propagators, the
full dots are $\hybb$ interaction vertices.
}
\label{diag:selfenergy}
\end{figure}

To determine the RG beta function $\beta(g)$ we evaluate the fermionic self-energy up to
one-loop order
\begin{equation}
\Sigma_{f_\sigma}(i\omega_n)=K g^2\frac{
\mu^{2\bar{r}}}{\beta}\sum_{i\omega_n^\prime}\int_{-\Lambda}^\Lambda d k\,|k|^r
\,\frac{1}{i\omega_n^\prime-k}\frac{1}{i\bar{\omega}_n-i\omega_n^\prime}
\end{equation}
corresponding to the diagram in Fig.~\ref{diag:selfenergy}a. Here, we have introduced the abbreviated
notation $i\bar{\omega}_n=i\omega_n-\lam$  for the Matsubara frequencies
$i\omega_n=i\pi(2n+1)/\beta$. In the  limit $\lam\to\infty$ and $\beta\to\infty$ we
obtain
\begin{eqnarray}
\Sigma_{f_\sigma}(i\omega_n)
&=& K g^2 \mu^{2\bar{r}}\int_0^\Lambda d \epsilon\,\epsilon^r \,\frac{1}{i\bar{\omega}_n-\epsilon}\\
&\approx&-Kg^2\left(\mu^{2\bar{r}}\frac{\Lambda^r}{r}+i\bar{\omega}_n \frac{1}{2\bar{r}}\right).\label{selfenergy}
\end{eqnarray}
An analogous expression is found for the bosonic self energy
$\Sigma_{b_{\bar{\alpha}}}$ depicted in Fig.~\ref{diag:selfenergy}b.
%
The renormalization factors are determined such that they cancel the $\frac{1}{2\bar{r}}$
pole in the self-energies minimally and  render the inverse Green's function finite. We
thus find
\begin{eqnarray}
Z_f=1-K\frac{g^2}{2\bar{r}},\quad Z_b=1-N\frac{g^2}{2\bar{r}}.
\end{eqnarray}
The mass counter-terms are  given by the real parts of the self-energies
\begin{eqnarray}
\delta\lambda_f=Kg^2\mu^{2\bar{r}}\frac{\Lambda^r}{r},\quad \delta\lambda_b=N g^2\mu^{2\bar{r}}\frac{\Lambda^r}{r}.
\end{eqnarray}
To one-loop order, there is no vertex renormalization of $g$, hence we have $Z_g$=1 at this
order (note that a $g^3$ diagram does not exist due to the directed nature of the
propagators).
The beta function
\begin{eqnarray}
\beta(g)\equiv\mu\frac{d g}{d\mu}
\end{eqnarray}
can now be obtained by taking the logarithmic $\mu$ derivative of Eq.~(\ref{g_r}). Since
$\mu\frac{d\hybb}{d\mu}=0$ we can solve for $\beta(\hyb)$ and finally obtain
\begin{eqnarray}\label{beta}
\beta(g)&=&-\frac{1-r}{2}g+\frac{K+N}{2}g^3
\end{eqnarray}
to one-loop order. One can also consider the flow away from criticality, i.e. the flow of
the renormalized tuning parameter $\eps$ using $S^2$ insertions. The resulting
correlation lengths exponent is given in Eq.~(\ref{nuz_2cam}) below.

\subsection{$\rmax < r < 1$}

For $r<1$ the trivial fixed point $\hyb^\ast=0$ is unstable,
and the critical properties are instead controlled by an interacting
fixed point (labelled ACR) at
\begin{eqnarray}\label{fp}
{\hyb^\ast}^2&=& \frac{1-r}{K+N}
\end{eqnarray}
with anomalous field dimensions
\begin{eqnarray}
\eta_b &=&\beta(g)\left.\frac{d\ln Z_b}{d g}\right|_{g^\ast}= N {\hyb^\ast}^2 \\
\eta_f &=&\beta(g)\left.\frac{d\ln Z_f}{d g}\right|_{g^\ast}=K{\hyb^\ast}^2 .
\end{eqnarray}
The corresponding RG flow diagram is displayed in Fig.~\ref{fig:flow2cam}c.

ACR describes a quantum phase transition below its upper-critical dimension. As a
result, low-energy observables calculated at and near ACR will be fully universal, i.e., cutoff-independent, and hyperscaling is fulfilled.


\subsection{$r\geq1$}

For all $r\geq 1$, i.e., above the upper-critical dimension, the phase transition between
LM and LM$^{\prime}$ is now controlled by the non-interacting FImp fixed point at
$\hyb=\eps=0$. Hence, for all $r>1$ the phase transition is a level crossing with
perturbative corrections -- this results e.g.~in a jump of the order parameter
$T\chi_{\rm loc}$ (see below), i.e., the transition is formally of first order.
Consequently, hyperscaling is violated, and all observables will depend upon the
UV cutoff.

For the marginal case, $r=1$, we expect a logarithmic flow of the marginally irrelevant
hybridization $g$, characteristic of the behavior at the upper-critical dimension. The RG
beta function
\begin{eqnarray}
\beta(\hyb) = \frac{K+N}{2} \hyb^3
\end{eqnarray}
can be integrated (recall $\beta(\hyb) \equiv \frac{d \hyb}{d \ln \ell}$ where $\ell$
describes the reduction of the UV cutoff $\Lambda \to \ell\Lambda$) to give
\begin{eqnarray}
\label{hybr1}
\hyb^2(\ell) = \frac{\hybb^2}{1-(K+N)\hybb^2\ln\ell}
\end{eqnarray}
with $\hyb(\ell\!=\!1) = \hybb$.
This result can be used to determine logarithmic corrections to observables.


\subsection{Observables near criticality}
\label{sec:2camobs}

\subsubsection{Correlation-length exponent}

We start with the correlation-length exponent, $\nu$, of the ACR fixed
point. This exponent describes the vanishing of the characteristic crossover temperature
in the vicinity of the critical point $T^\ast\propto(\varepsilon-\varepsilon^\ast)^\nu$.
The lowest-order result for $\nu$, which can be obtained either using the field-theoretic
RG scheme via composite operator insertions or using the familiar momentan shell scheme, is
\begin{equation}
\frac{1}{\nu} = r + {\cal O}(\bar{r}^2) ~~~(r<1).
\label{nuz_2cam}
\end{equation}

For $r\geq 1$ the transition is a level crossing, formally $\nu=1$.

\subsubsection{Local susceptibility}

\begin{figure}
\epsfxsize=3.2in
\centerline{\epsffile{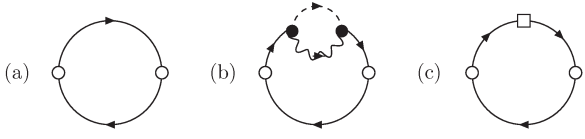}}
\caption{
Feynman diagrams entering $\langle \chi_{\rm loc} \rangle_{\lam}$ up to quadratic order
in $\hybb$. Open circles are sources,  and the blank boxes
denotes the counter terms $\delta\lambda_f$. Notation otherwise as in Fig.~\ref{diag:selfenergy}
}
\label{diag:lchi}
\end{figure}

\begin{figure}
\epsfxsize=3.2in
\centerline{\epsffile{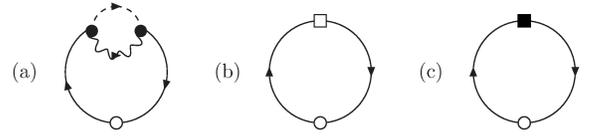}}
\caption{
Feynman diagrams entering the corrections to the unperturbed part of $\langle
\hat{Q}\rangle_{\lam}$ up to quadratic order in $\hybb$. Notation as in the previous
figures. Black boxes denote the counter terms $\delta\lambda_b$.
}
\label{diag:Q}
\end{figure}

The local susceptibility $\chi_{\rm loc}=\chi_{\rm imp,imp}$ at the critical point
follows the scaling behavior $\chi_{\rm loc}\propto T^{-1+\eta_{\chi}}$ with an anomalous
exponent $\eta_{\rm \chi}$. To obtain the corrections to the tree-level result $\chi_{\rm
loc} \propto T^{-1}$ we introduce a $\chi_{\rm{loc}}$ renormalization factor $Z_{\chi}$
from which  one obtains the anomalous exponent according to
\begin{equation}
\label{etachidef2}
\eta_\chi = \beta(\hyb) \left.\frac{d \ln Z_\chi} {d \hyb} \right|_{\hyb^\ast}.
\end{equation}
We determine $Z_\chi$ by calculating $\langle \chi_{\rm loc}\rangle$ directly using
perturbative corrections up to quadratic order in $\hybb$. The corresponding diagrams
entering $\langle \chi_{\rm loc}\rangle_{\lam}$  are given in Fig.~\ref{diag:lchi}.
In terms of the renormalized coupling constant $\hyb$  we find at the energy scale $\mu=T$
\begin{eqnarray}
\langle \chi_{\rm loc} \rangle_{\lam} &=& e^{-\frac{\lam}{T}}\left(\frac{1}{2T} + \frac{K \hyb^2}{T} \int_0^{\frac{\Lambda}{T}}dx \frac{x^r}{x^3}\left[2 \tanh\frac{x}{2} \right.\right. \nonumber \\
 &&\left.\left. +\frac{x^2}{2}\tanh{\frac{x}{2}}-x-\frac{x^2}{2}\right]\right).
\end{eqnarray}
Furthermore, the denominator $\langle\hat{Q}\rangle_{\lam}$ receives corrections from the
diagrams in Fig.~\ref{diag:Q}, resulting in
\begin{eqnarray}\label{cor:Q}
\langle\hat{Q}\rangle_{\lam}&=&(N+K)e^{-\frac{\lam}{T}}\\
&& +2NK\hyb^2 e^{-\frac{\lam}{T}}\int_0^{\frac{\Lambda}{T}}dx \frac{x^r}{x}\left[\tanh\frac{x}{2}-1\right].\nonumber
\end{eqnarray}
The local susceptibility $\chi_{\rm loc}$ can  then be directly obtained by
Eq.~(\ref{observables_lambda}). The renormalization factor $Z_\chi$ is then determined,
using minimal subtraction of poles, in an expansion in $\bar{r}$ as
\begin{eqnarray}
Z_\chi=1-K\hyb^2 \frac{1}{\bar{r}}
\end{eqnarray}
and from this we can directly deduce the anomalous exponent of the local spin
suceptibility
\begin{eqnarray}
\eta_{\chi}^{(\rm spin)} = 2 K {g^\ast}^2 = \frac{2 K}{K+N} (1-r).
\end{eqnarray}
The expression for $\eta_{\chi}^{(\rm flavor)}$ follows by the replacement $K
\leftrightarrow N$.

Above the upper-critical dimension, $r>1$, we simply have $\eta_\chi =0$ and, thus,
$\chi_{\rm loc}^{(\rm spin)}, \chi_{\rm loc}^{(\rm flavor)} \propto T^{-1}$ or $\propto \omega^{-1}$.
For the marginal case $r=1$, a calculation analogous to that in Ref.~\onlinecite{FV04},
using Eq.~\eqref{hybr1}, gives
\begin{eqnarray}
\chi_{\rm loc}^{(\rm spin)} \propto \frac{1}{\w|\ln\w|^{2K/(K+N)}}~~~(r=1).
\end{eqnarray}

\subsubsection{Order parameter}

Inside the stable phases LM and LM$^\prime$, $\chi_{\rm loc}$ can be used to define an
order parameter for the quantum phase transition:
$T\chi_{\rm loc}^{\rm spin}$ is finite (zero) for $\epsb<0$ ($\epsb>0$), similarly
$T\chi_{\rm loc}^{(\rm flavor)}$ is finite (zero) for $\epsb>0$ ($\epsb<0$). Approaching
the critical point, both order parameters vanish continuously according to
\begin{eqnarray}
T\chi_{\rm loc}^{(\rm spin)}   &\propto& (-\epsb)^{\nu\eta_\chi^{(\rm spin)}}\,,\nonumber\\
T\chi_{\rm loc}^{(\rm flavor)} &\propto& \epsb^{\nu\eta_\chi^{(\rm flavor)}}
\end{eqnarray}
for $r<1$, which follows e.g. from hyperscaling. Note that these order parameters display
a jump upon crossing the transition for $r>1$.

\subsubsection{Impurity susceptibility}

\begin{figure}
\epsfxsize=3.2in
\centerline{\epsffile{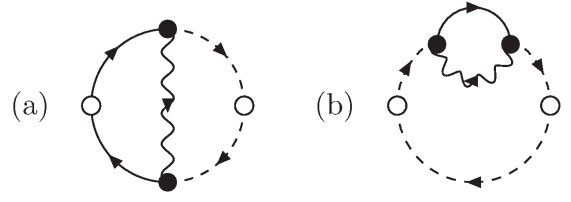}}
\caption{
Further Feynman diagrams entering corrections to the impurity susceptibility to
quadratic order in $\hybb$. Notation as in the previous figures.
}
\label{diag:chi_imp}
\end{figure}

The evaluation of the impurity susceptibility $\chi_{\rm imp}$  to second  order in
$\hybb$ requires the summation of further diagrams as depicted in
Fig.~\ref{diag:chi_imp}. In terms of the renormalized coupling $\hyb$ we obtain
\begin{eqnarray}
2\langle \chi_{\rm b,imp}\rangle_{\lam}&=&\frac{K\hyb^2}{T} e^{-\frac{\lam}{T}} \int_0^{\frac{\Lambda}{T}}dx \frac{x^r}{x^3}\Big[x+\frac{x}{\cosh^2{\frac{x}{2}}} \nonumber \\
&& -4\tanh{\frac{x}{2 }}\Big],
\end{eqnarray}
and
\begin{eqnarray}
\langle \chi_{\rm b,b}\rangle_{\lam}- \langle\chi_{\rm b,b}^{0}\rangle_{\lam}&=&\frac{K\hyb^2}{T} e^{-\frac{\lam}{T}}\int_0^{\frac{\Lambda}{T}}dx \frac{x^r}{x^3}\Big[2\tanh\frac{x}{2} \nonumber \\
 &&   - \frac{x}{\cosh^2{\frac{x}{2}}}-\frac{ x^2 \tanh{\frac{x}{2}}}{2 \cosh^2{\frac{x}{2}}} \Big].
\end{eqnarray}
Collecting all contributions to $\chi_{\rm imp}$ to second order in $\hyb$ the poles
present in the $\chi_{\rm loc}$ diagrams cancel and the remaining momentum integrals are
UV convergent for $r<1$. Performing these integrals for $r<1$ the impurity susceptibility
reads
\begin{eqnarray}
T \chi_{\rm imp}^{(\rm spin)}
&=& \frac{1}{2(N\!+\!K)} \left[1- \hyb^2 K\left(1+\ln4-\frac{2N}{N\!+\!K}\ln4\right)\right]\nonumber \\
 & &+ {\cal O}(\hyb^4).
\end{eqnarray}
As above, the expression for $T\chi_{\rm imp}^{(\rm flavor)}$ follows by the replacement $K
\leftrightarrow N$.
With the value of the coupling at the ACR fixed point (\ref{fp})
we finally find for $N=K=2$, to leading order in $(1-r)$,
\begin{equation}
\label{tchi_2cam}
T\chi_{\rm imp}^{(\rm spin)} = \left\{
\begin{array}{ll}
\frac{1}{8} -\frac{1}{16}(1-r) + {\cal O}(\bar{r}^2)   & (r < 1) \\[2mm]
\frac{1}{8}                                         & (r \geq 1) \\
\end{array}
\right.
\,,
\end{equation}
with $T\chi_{\rm imp}^{(\rm flavor)} = T\chi_{\rm imp}^{(\rm spin)}$ due to the
emergent $Z_2$ symmetry \eqref{ph2}. A comparison to NRG data is in Fig.~\ref{fig:tchi}.
Note that $T\chi_{\rm imp}$ receives only weak additive logarithmic corrections at $r=1$;
multiplicative logs as in $\chi_{\rm loc}$ are absent here. The same applies to $S_{\rm imp}$
below.

\subsubsection{Impurity entropy}
\label{sec:2camentr}

The impurity contribution to the entropy can be obtained from the thermodynamic potential
$\Omega_{\rm imp}$  by $S_{\rm imp} =-\partial_T \Omega_{\rm imp}$. At the FImp fixed
point the entropy is $S_{\rm imp} = \ln(N+K)$, and the lowest-order correction is
computed by expanding the thermodynamic potential in the renormalized hybridization
$\hyb$. Note that this correction vanishes for $r\geq 1$, as $\hyb^\ast=0$ there. Here,
we write the partition function in the physical sector of the Hilbert space ($\hat{Q}=1$)
as\cite{MKMV}
\begin{eqnarray}
\frac{Z_{\rm{imp}}}{Z_{\rm{imp},0}}=\lim_{\lambda_0\rightarrow \infty} \frac{\langle\hat{Q}\rangle_{\lambda_0}}{\langle\hat{Q}\rangle_{\lambda_0,0}},
\end{eqnarray}
where $\langle\dots\rangle_{\lambda_0,0}$ is the expectation value in the presence of $\lambda_0$ without coupling to the bath.
The thermodynamic potential is given by
\begin{eqnarray}\label{therm_potential}
\Omega_{\rm imp}-\Omega_{\rm imp,0}=-T\ln\frac{Z_{\rm{imp}}}{Z_{\rm{imp},0}}.
\end{eqnarray}
The correction to $\langle\hat{Q}\rangle_{\lambda_0}$ due to the coupling to the bath up to quadratic order in  $\hyb$ has already been calculated in Eq.~(\ref{cor:Q}) which enables us now to directly evaluate Eq.~(\ref{therm_potential}).
Taking the temperature derivative of the resulting expression
and evaluating the remaining integral in the
limit $T\to 0$ and $r\to 1$ we obtain
\begin{eqnarray}
\label{simp3}
S_{\rm imp} = \ln(K+N) - \hyb^2\,\frac{4NK}{N+K}\ln 2  + {\cal O}(\hyb^4).
\end{eqnarray}
As expected, the entropy correction is fully universal and finite in the limit $T\to 0$.
Note that in higher-order terms of the diagrammatic expansion for the thermodynamic
potential $\Omega_{\rm{imp}}$ disconnected diagrams appear.\cite{MKMV}

Finally, inserting the fixed-point value of the coupling $\hyb$ into Eq.~(\ref{simp3}),
we find the impurity entropy in the two-channel case $N=2$ and $K=2$ to be
\begin{equation}
\label{simp_2cam}
S_{\rm imp} =
\left\{
\begin{array}{ll}
\ln 4 - (1-r)\ln 2  + {\cal O}(\bar{r}^2)   & (r < 1) \\
\ln 4                                       & (r \geq 1) \\
\end{array}
\right.
\,.
\end{equation}
Comparison with the numerical result again shows very good agreement, Fig.~\ref{fig:simp}.

\subsubsection{Conduction-electron $T$ matrix}

For the single-channel pseudogap Kondo problem, it has been shown that the
conduction-electron $T$ matrix $T(\w)$ displays a power-law divergence of the form
$T(\w) \propto |\w|^{-r}$ at all intermediate-coupling fixed points.\cite{VF04,FV04}
Analytically, this follows -- for all perturbative expansions -- from the diagrammatic
structure of the $T$ matrix (or, alternatively, from a Ward identity).

In the two-channel case, we find the same qualitative arguments to apply, i.e., at the
NFL, SCR, and ACR fixed points the $T$ matrix obeys the exact result
\begin{equation}
T(\w) \propto |\w|^{-r} ~~~(r<1).
\end{equation}
At one-loop level, an explicit calculation gives $\eta_T = (K+N)\hyb^2$ which yields $\eta_T=
1-r$ as expected. For $r>1$, ${\rm Im}\,T(\w) \propto \delta(\w)$.

For $r=1$, the logarithmic flow of the coupling, Eq.~\eqref{hybr1}, can used to deduce
$T(\w) \propto 1/(\w|\ln\w|)$ which gives
\begin{equation}
{\rm Im}\,T(\w) \propto \frac{1}{\w|\ln\w|^2}
\end{equation}
in analogy to Ref.~\onlinecite{FV04}.


\section{Conclusions}

We have explored the two-channel Kondo effect for magnetic impurities embedded into a
fermionic host with a power-law pseudogapped density of states. We have determined the
phase diagram as function of the DOS exponent $r$ and discussed the boundary quantum
phase transitions of the relevant Kondo and Anderson models. These transition are
described by fermionic (as opposed to usual bosonic) quantum field theories; from their
properties we conclude that there is no underlying CFT description.

Our results demonstrate the versatility of the Anderson-model epsilon expansions
developed in Refs.~\onlinecite{VF04,FV04}: Those have not only allowed a full
understanding of the critical behavior of the single-channel ($S=1/2$) pseudogap Kondo
problem, but also of the corresponding underscreened\cite{FV05} and overscreened
pseudogap Kondo models (this work). Further applications, e.g., to multi-impurity models,
appear possible. Also, the Anderson-model formulation should enable studies of
non-equilibrium dynamics of pseudogap Kondo problems.

Our results are of potential relevance to two-channel impurities in unconventional
superconductors and in graphene; for the latter case the extension of the present
calculations to finite bias\cite{epl} is an interesting future topic.


\acknowledgments

We thank A.~J.~Schofield and, in particular, K.~Ingersent for helpful discussions. This
research has been supported by the Deutsche Forschungsgemeinschaft through SFB 608 (LF),
FOR 960 (AB, MV), and AN 275/6-2 (FBA). FBA and MV also acknowledge support from the
German-Israeli-Foundation.


\appendix


\section{Compactified $\sigma$-$\tau$ Kondo model and O(3)-symmetric Anderson model}
\label{sigmatau}

Here we briefly discuss an alternative formulation of the two-channel Kondo model which
eventually leads to a theory of non-interacting Majorana fermions.

\subsection{Metallic bath, $r$=0}

The low-energy physics of the standard two-channel Kondo problem ($r=0$) has been argued
to be equivalent to that of the so-called $\sigma$-$\tau$ Kondo model -- this is a
``compactified'' single-channel Kondo model where the roles of the two screening channels
are taken by spin and a charge pseudospin.\cite{schofield1,schofield2} The corresponding
Hamiltonian can be expressed as
\begin{eqnarray}
\mathcal{H}_{\sigma\mbox{-}\tau} &=& \left[J_1 \vec{\sigma}(0)+J_2\vec{\tau}(0)\right]\cdot \vec{S}
+ \int_{-\Lambda}^{\Lambda} dk \,
  k \, c_{k \sigma}^\dagger c_{k \sigma} ,
\end{eqnarray}
where, as above, $\vec{S}$ is a spin-1/2 SU(2) spin and we  have represented the bath by linearly dispersing chiral
fermions $c_{k \sigma}$. Spin degrees of freedom $\sigma=\uparrow, \downarrow$ are implicitly summed.
The conduction electron spin $\vec{\sigma}(0)$ and  pseudospin $\vec{\tau}(0)$
are defined as
\begin{eqnarray}
\vec{\sigma}(0)&=&(c_\uparrow^\dagger(0),c_\downarrow^\dagger(0))\cdot\vec{\tau}\cdot \begin{pmatrix} c_{\uparrow}(0)\\c_{\downarrow}(0)\end{pmatrix}\\
\vec{\tau}(0)&=&(c_\uparrow^\dagger(0),c_\downarrow(0))\cdot\vec{\tau}\cdot \begin{pmatrix} c_{\uparrow}(0)\\c_{\downarrow}^\dagger(0)\end{pmatrix},
\end{eqnarray}
where $c_{\sigma}(0) = \int d k c_{k \sigma}$.
Interestingly, the low-energy physics of the $\sigma$-$\tau$ Kondo model is described by
a fixed point with non-Fermi liquid behavior which is located at {\em strong} coupling,
not at intermediate coupling as in the two-channel Kondo problem. The equivalence of the
two-channel Kondo model and the $\sigma$-$\tau$ Kondo model has been established using
bosonization and conformal field theory techniques. \cite{schofield1,schofield2,bulla97}

The nature of the low-energy non-Fermi liquid becomes transparent by considering the
so-called O(3)-symmetric Anderson model which displays an anomalous hybridization term.
Its Hamiltonian is given by $\mathcal{H}_{\rm O(3)}=\mathcal{H}_{\rm
1cA}+\mathcal{H}_{\rm ahyb}$ with
\begin{eqnarray}
\label{symmetricA}
\mathcal{H}_{\rm 1cA} &=&
U \left( n_{f\uparrow}-\frac{1}{2}\right)\left(n_{f\downarrow}-\frac{1}{2}\right) \\
& &+ \int_{-\Lambda}^{\Lambda} dk \,
  k \, c_{k \sigma}^\dagger c_{k \sigma}
 + \hybb \sum_\sigma\left[f^\dagger_\sigma c_{\sigma}(0) + \rm{H.c.} \right], \nonumber \\
\mathcal{H}_{\rm ahyb}&=&-g_a \left[f^\dagger_\downarrow c^\dagger_{\downarrow}(0) +f_\downarrow^\dagger c_{\downarrow}(0) +\rm{H.c.} \right],
\end{eqnarray}
where $f^\dagger_\sigma$ creates the localized impurity state with spin $\sigma$ and
$n_{f\sigma}=f_\sigma^\dagger f_\sigma$. Note that the chemical potential on the impurity
site  has been  chosen  such that the model is p-h symmetric. In the Kondo
limit, the  O(3)-symmetric Anderson  model maps onto the $\sigma$-$\tau$ Kondo model.\cite{bulla97}

Appealing to the adiabatic continuity between the $U=0$ and large-$U$ limits in this
Anderson model suggests to discuss the weakly-interacting case.
The Hamiltonian $\mathcal{H}_{\rm O(3)}$ can be conveniently re-written in terms of Majorana fermions:
\begin{eqnarray}\label{H_majorana}
\mathcal{H}_{\rm O(3)} &=& U d_1d_2d_3d_0  +i \sum_{\alpha=0}^3\int_{-\Lambda}^{\Lambda}dk  \,k \,\psi_{-k\alpha} \psi_{k\alpha}\\ & & +i\hybb\sum_{\alpha=1}^3 \psi_\alpha(0)d_\alpha   +i (\hybb-2g_a) \psi_0(0)d_0
.\nonumber
\end{eqnarray}
Here, the impurity Majorana fermions are defined by
\begin{eqnarray}
f_\uparrow&=&\frac{1}{\sqrt{2}}(d_1-i d_2),\quad f_\downarrow=\frac{1}{\sqrt{2}}(-d_3+i d_0),
\end{eqnarray}
where $d^\dagger_\alpha=d_\alpha$ and $\{d_\alpha,d_\beta\}=\delta_{\alpha,\beta}$ for $\alpha=0,1,2,3$.
The Majorana fermions for the conduction electrons are defined similarly\cite{bulla97} which in Fourier space reads
\begin{eqnarray}
c_{k\uparrow}= \frac{1}{\sqrt{2}}\left(-i\psi_{k1}-\psi_{k2}\right),\quad
c_{k\downarrow}=\frac{1}{\sqrt{2}}\left(i\psi_{k3}-\psi_{k4}\right)\nonumber\\
\end{eqnarray}
with $\psi^\dagger_{k\alpha}=\psi_{-k\alpha}$. Remarkably, for
$2g_a=\hybb$ the impurity couples only via $d_\alpha$ for $\alpha=1,2,3$ to the
conduction Majorana fermions while $d_0$ remains free.

The Hamiltonian in Eq.~(\ref{H_majorana})  is in particular suitable to study
thermodynamic quantities. Note that for $U=0$ the model is exactly solvable and the impurity Green's functions $G_\alpha(\tau)=-\langle T_\tau d_\alpha(\tau)d_\alpha(0)\rangle$ are known exactly. For $U=0$ and $2g_a=\hybb$, their Fourier counterparts read
\begin{eqnarray}\label{impGreens}
G_0(i\omega_n)=\frac{1}{i\omega_n},\quad G_\alpha(i\omega_n)=\frac{1}{i\omega_n+i A_0 {\rm sgn}(\omega_n)},
\end{eqnarray}
for $\alpha=1,2,3$. Here, we have introduced $A_0=\pi \hybb^2$.

A straightforward calculation now shows then that there is a
residual impurity entropy $S_{\rm imp}=\frac{1}{2}\ln 2$ of a free Majorana fermion.\cite{zhang1996}
By adiabatic continuity, this entropy persists into the regime of large $U$ and then
corresponds to the entropy of the overscreened two-channel Kondo impurity.

\subsection{Pseudogap bath}

The obvious question is whether the $\sigma$-$\tau$ Kondo and O(3)-symmetric Anderson
models continue to represent the physics of the two-channel Kondo problem for a pseudogap
bath DOS with $r>0$. To answer this, let us consider the non-interacting limit and $2g_a=\hybb$ of the
O(3) Anderson model. The Green's functions in Eq.~(\ref{impGreens}) for $\omega_n/\Lambda\ll 1$ are now given by \cite{FV04}
\begin{eqnarray}
G_0(i\omega_n)&=&\frac{1}{i\omega_n},\\
G_\alpha(i\omega_n)&=&\frac{1}{i\omega_n+i A_0{\rm sgn}(\omega_n)|\omega_n|^r}
\end{eqnarray}
for $\alpha=1,2,3$.
They yield an impurity entropy of
\begin{eqnarray}
S_{\rm imp} =\frac{1}{2}\ln2 + \frac{3}{2}r\ln2.
\end{eqnarray}
and an impurity susceptibility of
\begin{eqnarray}
T\chi_{\rm imp}(T) =\frac{3r}{32}.
\end{eqnarray}
This result is not in agreement with the numerical data in Figs.~\ref{fig:tchi} and
\ref{fig:simp}, which instead are well fitted by $S_{\rm imp} =\frac{1}{2}\ln2 + 2r\ln2$
and  $T\chi_{\rm imp}(T)=\frac{r}{6}$ (Ref.~\onlinecite{GBI}). We are forced to conclude
that the low-energy behavior of the $\sigma$-$\tau$ Kondo and O(3)-symmetric Anderson
models is {\em not} identical to that of the two-channel Kondo model once $r>0$. In other
words, the equivalence is restricted to the metallic $r=0$ case. Given the fact that
neither bosonization nor CFT appear to be applicable to the pseudogap Kondo models, this
may not come as a surprise.

Let us finish
with the remark that an extension of the Majorana resonant-level model that corresponds
to the solvable point of the two-channel Kondo model\cite{emery1992} to a pseudogap DOS
also does not yield the numerically found impurity entropy.



\begin{thebibliography}{99}
\bibitem{asc_note}
The fixed point LM$^\prime$ was labelled ASC (for asymmetric strong coupling) in
Ref.~\onlinecite{GBI}.
\end{thebibliography}
\end{document}